\documentclass[conference]{IEEEtran}
\IEEEoverridecommandlockouts
\usepackage{cite}
\usepackage{multirow}
\usepackage{amsmath,amssymb,amsfonts}
\usepackage{algorithmic}
\usepackage{graphicx}
\usepackage{textcomp}
\usepackage{xcolor}
\def\BibTeX{{\rm B\kern-.05em{\sc i\kern-.025em b}\kern-.08em
    T\kern-.1667em\lower.7ex\hbox{E}\kern-.125emX}}
\begin{document}

\title{Scalable Machine Learning Architecture for Neonatal Seizure Detection on Ultra-Edge Devices\\}

\author{\IEEEauthorblockN{Vishal Nagarajan}
\IEEEauthorblockA{\textit{Department of Computer Science and Engineering} \\
\textit{Sri Sivasubramaniya Nadar College of Engineering}\\
Chennai, India \\
vishal18198@cse.ssn.edu.in}
\and
\IEEEauthorblockN{Ashwini Muralidharan}
\IEEEauthorblockA{\textit{Department of Electrical and Electronics Engineering} \\
\textit{Sri Sivasubramaniya Nadar College of Engineering}\\
Chennai, India \\
ashwini183001014@eee.ssn.edu.in}
\and
\IEEEauthorblockN{Deekshitha Sriraman}
\IEEEauthorblockA{\textit{Department of Electrical and Electronics Engineering} \\
\textit{Sri Sivasubramaniya Nadar College of Engineering}\\
Chennai, India\\
deekshitha183001019@eee.ssn.edu.in}
\and
\IEEEauthorblockN{Pravin Kumar S}
\IEEEauthorblockA{\textit{Department of Biomedical Engineering} \\
\textit{Sri Sivasubramaniya Nadar College of Engineering}\\
Chennai, India \\
pravinkumars@ssn.edu.in}
}

\maketitle

\begin{abstract}

Neonatal seizures are a commonly encountered neurological condition. They are the first clinical signs of a serious neurological disorder. Thus, rapid recognition and treatment are necessary to prevent serious fatalities. The use of electroencephalography (EEG) in the field of neurology allows precise diagnosis of several medical conditions. However, interpreting EEG signals needs the attention of highly specialized staff since the infant brain is developmentally immature during the neonatal period. Detecting seizures on time could potentially prevent the negative effects on the neurocognitive development of the infants. In recent years, neonatal seizure detection using machine learning algorithms have been gaining traction. Since there is a need for the classification of bio-signals to be computationally inexpensive in the case of seizure detection, this research presents a machine learning (ML) based architecture that operates with comparable predictive performance as previous models but with minimum level configuration. The proposed classifier was trained and tested on a public dataset of NICU seizures recorded at the Helsinki University Hospital. Our architecture achieved a best sensitivity of 87\%, which is 6\% more than that of the standard ML model chosen in this study. The model size of the ML classifier is optimized to just 4.84 KB with minimum prediction time of 182.61 milliseconds, thus enabling it to be deployed on wearable ultra-edge devices for quick and accurate response and obviating the need for cloud-based and other such exhaustive computational methods.

\end{abstract}


\section{Introduction}
\label{Introduction}
Seizures are excessive, synchronous neuronal activity in the brain. Neonatal seizures are those that occur from birth to the end of the neonatal period (first 28 days of life of a full-term infant). The infant brain is highly susceptible to seizures, and it is categorized as one of the most severe complications in the Neonatal Intensive Care Unit (NICU), owing to its several adverse consequences. Firstly, due to its association with increased brain damage, it stunts neuro-developmental and intellectual growth of the neonate. Prolonged and frequent seizures are associated with more severe effects on cognitive functioning of the developing brain \cite{1}. Secondly, neonatal seizures often go undetected or unaddressed. Untreated seizures not only affect brain development, but in 15\% - 18\% cases \cite{2}, it can lead to death. Moreover, the symptoms of neonatal seizures are considerably similar to normal neonatal behavior and do not have any observable clinical manifestations. Its clinical subtlety and inconspicuous nature make it difficult to diagnose, thus delaying necessary treatment. Although there are many studies conducted in the field of epileptic seizure detection for adults, those results and findings are not applicable for neonatal seizure detection. Due to the disparate nature between the two seizures \cite{3}, there is a pressing need for neonatal seizure detection. So far, the most common way of detecting neonatal seizures remains the visual interpretation of electroencephalogram (EEG) signals along with a clinical neurophysiologist to identify and classify the abnormality \cite{3}.  This method is highly time-consuming and requires the expertise of medical personnel, ultimately delaying treatment. Therefore, this research aims at developing a neonatal seizure detection architecture that is feasible.

This paper is structured as follows. Section \ref{Related Work} contains the literature of existing work in seizure detection. Section \ref{Dataset} briefly describes the dataset used. Section \ref{Dataset Preprocessing} elucidates on the preprocessing procedure. Section \ref{Pipeline Architecture} explains the architecture in detail. Section \ref{Results and Discussion} discusses and analyzes the obtained results.

\section{Related Work}
\label{Related Work}

For neonatal seizure detection (NSD), there exist several machine learning and deep learning approaches. Pavel et al. \cite{4} developed an automated seizure detection algorithm called ANSeR and reported sensitivity - 81.3\% and 89.5\%, and specificity - 84.4\% and 89.1\% in the algorithm and non-algorithm group respectively. Temko et al. \cite{5} proposed an SVM-based NSD system with a Good Detection Rate (GDR) of 89\% and 1 false seizure detection per hour. Temko et al. \cite{6} and Tapani et al. \cite{7} proposed machine learning algorithms which showcased Area Under receiver-operating-characteristic Curve (AUC) of 97\% and 98\% respectively. They remain as the state-of-art (SoA) approaches for NSD using machine learning techniques. But the clinical feasibility of their approach is unwarranted as there is no hardware implementation to back the results. 

In order to circumvent the process of feature extraction, extensive research has been done using deep learning. Frassineti et al. \cite{8} proposed a hybrid system that combined techniques related to Stationary Wavelet Transform (SWT) and Deep Neural Networks including Fully Convolutional Networks. Their proposed method was validated on a public dataset of NICU seizures recorded at the Helsinki University Hospital, Finland and achieved AUC of 81\% and GDR of 77\%. With the main objective of automatically optimizing feature selection and classification, Ansari et al. \cite{9} proposed a model that uses Convolutional Neural Networks and Random Forest. Their proposed model delivered AUC of 83\% for the total database.

Lasefr et al. \cite{10} had developed a smartphone application that monitored EEG signals for epileptic seizures which runs on cloud servers. However, this approach is unsafe as it leaves a backdoor for potential privacy breach of users’ health data. Further, high operational latencies due to two-way data transmission to and from the cloud server severely affects the response time and thus, is not ideal for real-time monitoring. 

Our research proposes and implements an on-device, real-time detection pipeline that is clinically feasible for NSD. The proposed machine learning classifier detects seizures while consuming no more than 4.8 KB RAM and processing time of a few milliseconds.

\section{Dataset}
\label{Dataset}

A dataset \cite{11} composed of 18 channel EEG measurements recorded at the NICU of the Children’s Hospital, Helsinki University Central Hospital, Finland were used for training and testing of the algorithm. Seventy-nine neonates participated in the study, approved by the Institutional Ethics Committee of the Helsinki University Hospital, Finland and the recorded data was sampled at 256 Hz. For the recordings, 19 electrodes were placed according to the international 10-20 system with a bipolar montage. Three experts annotated the EEG data for the presence of seizures. An average of 460 seizures were annotated per expert in the dataset.  39 neonates had seizures by consensus and 22 were seizure free by consensus. The data is available publicly as 79 \textit{.edf} files along with annotations by the three experts as \textit{.mat} files. After extraction, 3 out of the 79 files were discovered to be corrupted, hence discarded from the study. This dataset is dubbed as the \textit{Helsinki dataset} and is thoroughly described in \cite{7}. 

\section{Dataset Preprocessing}
\label{Dataset Preprocessing}

\begin{figure}[htbp]
\centerline{\includegraphics[scale=0.45]{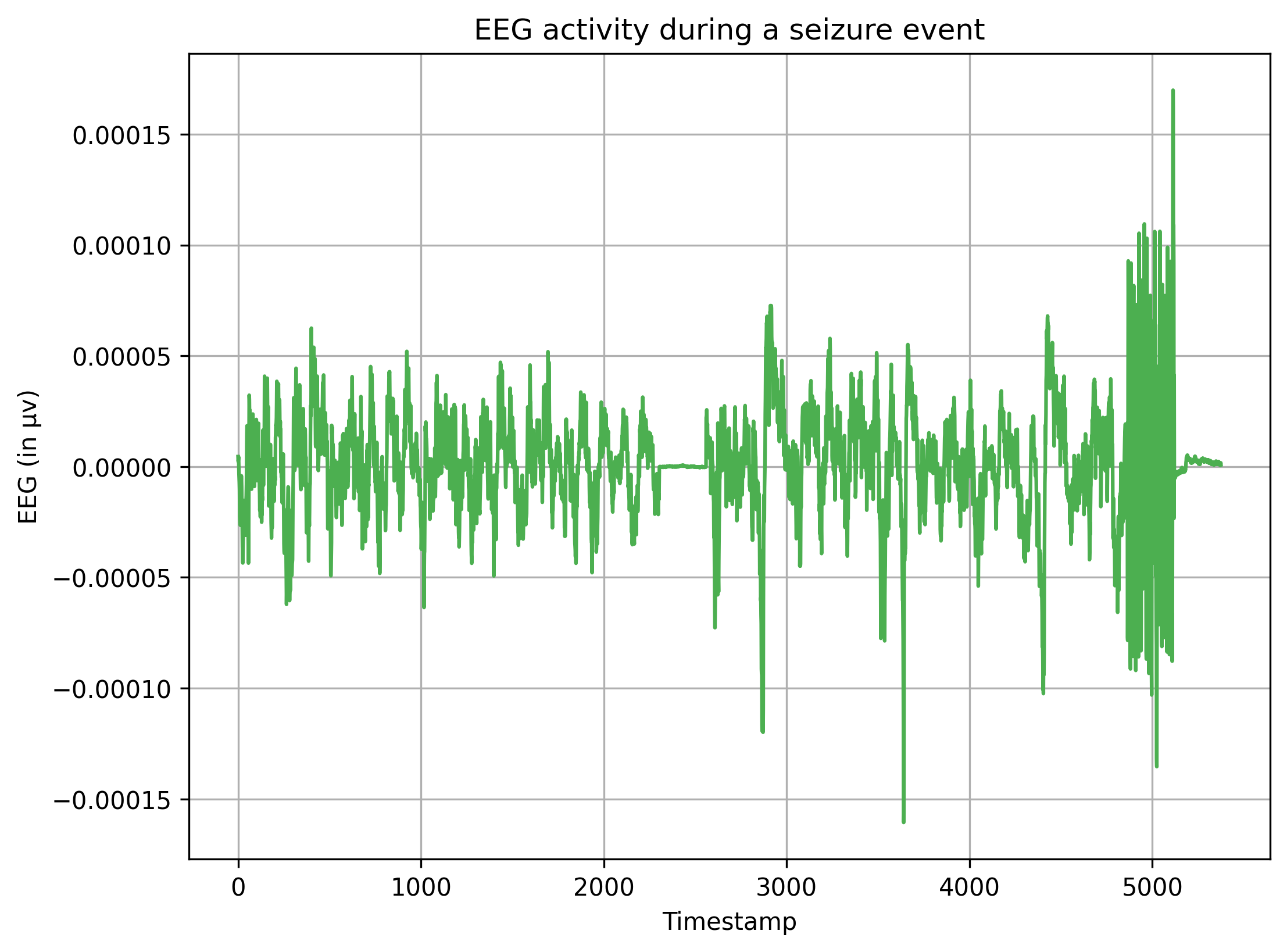}}
\caption{EEG activity during seizure event}
\label{seizure_activity}
\end{figure}

\begin{figure}[htbp]
\centerline{\includegraphics[scale=0.55]{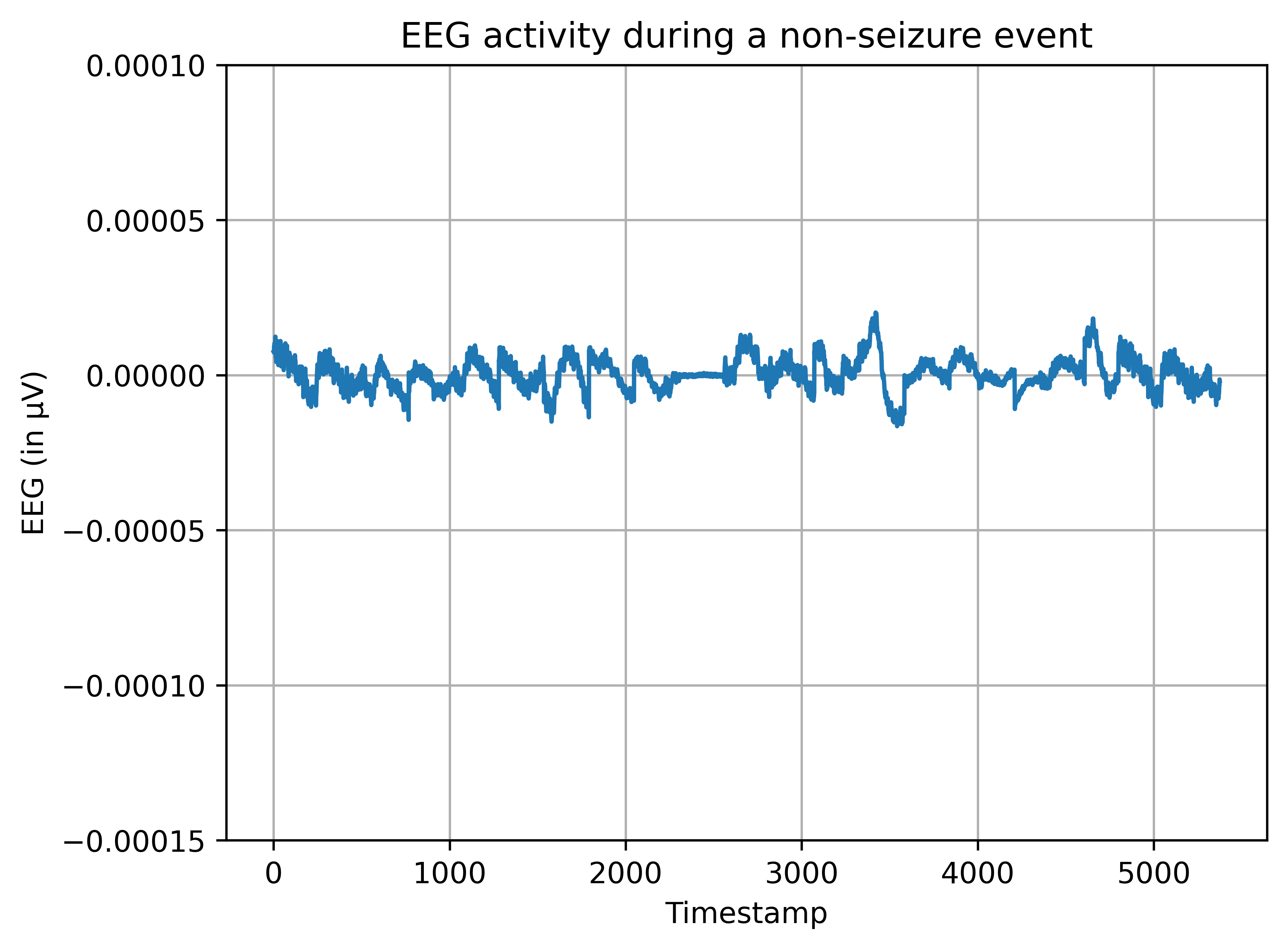}}
\caption{EEG activity under normal conditions}
\label{non_seizure_activity}
\end{figure}

The EEG data was first downsampled from 256 Hz to 32 Hz \cite{12, 13}. The raw data was filtered using a high-pass filter with a cut-off frequency of 0.5 Hz. The data was further rescaled to values between 0 and 1 by applying min-max scaler. The data from each channel was then segmented into various window lengths of 1, 2, 4, 8 and 16 seconds in this study \cite{10}. For each window length, it was stored as a 1-dimensional window of $w \times f_s$ time-steps, where $w$ is the length of the window in seconds and $f_s$ is the sampling frequency in Hz.

\begin{figure}
\centering
\includegraphics[scale = 0.5]{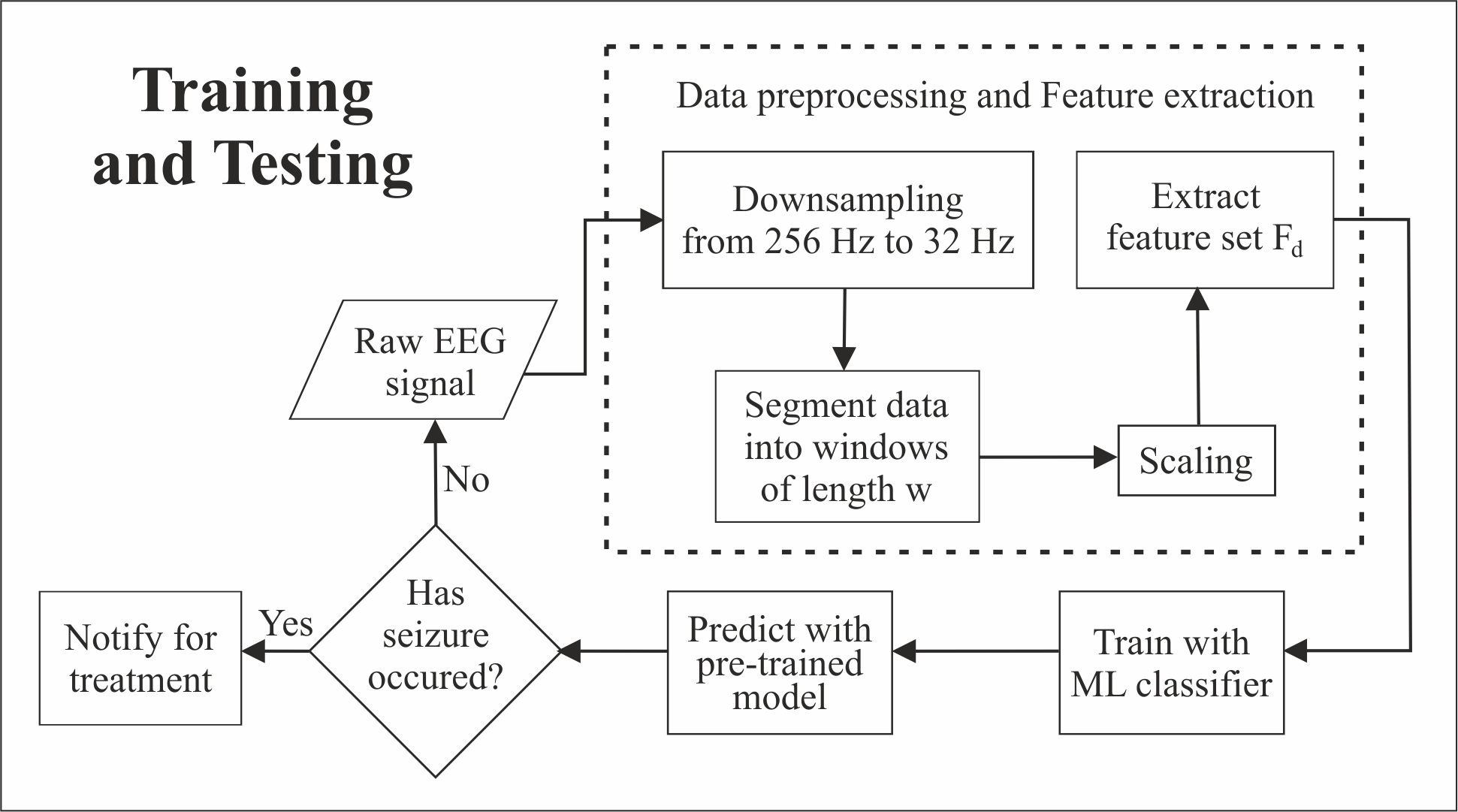}
\caption{Pipeline architecture}
\label{training_and_testing}
\end{figure}

\vspace{-2mm}

\begin{equation}
\label{eq:1}
    \delta = [\delta_1 \quad \delta_2 \quad ... \quad \delta_{w\times{f_s}}]_{w\times{f_s}}
\end{equation}

The multi-channel output annotations were transformed into a single decision value of either $1$ (indicating seizure) or $0$ (for non-seizure) by comparing the number of seizure annotations in the window. If the number of seizure annotations exceeded the threshold value $t$, then the window was labelled $1$. Fig. \ref{seizure_activity} and Fig. \ref{non_seizure_activity} compare EEG signal activity during ictal and normal conditions for one segment of data. Fig. \ref{seizure_activity} represents increased signal activity during the occurrence of seizure. Fig. \ref{non_seizure_activity} represents activity of signal under regular conditions. 

\section{Pipeline Architecture}
\label{Pipeline Architecture}

This end-to-end architecture receives raw EEG signal, processes it and classifies it as ictal or normal activity. After preprocessing, the signal is passed to a feature extraction engine that extracts the necessary feature set $F_d$. It is followed by a scalable machine learning (ML) classifier that performs prediction. The architecture is illustrated in Fig. \ref{training_and_testing}. It is further explained in-depth in the following subsections.

\subsection{Feature Extractor}
\label{Feature Extractor}

Powerful feature extraction system is necessary as features provide useful and relative information that aids seizure detection algorithm to discriminate between seizure and non-seizure events. Each segment of  EEG window $\delta$ was converted to 11 human-engineered features per channel, as listed and defined in Table \ref{Feature Set}.

For this study, $7$ time domain features and $4$ entropy domain features were extracted as they were relevant owing to their continued appearance in previous literature \cite{7}. Features such as mean, standard deviation, kurtosis and skewness are statistical in nature and are frequently used to differentiate between ictal and normal patterns \cite{14}. Other features such as Shannon Entropy, Approximate Entropy, Sample Entropy and Permutation Entropy are entropy-based indicators which determine the uncertainties and complexities of decomposed signals. The Hjorth parameters of activity, mobility and complexity are statistical properties in the time domain, commonly used in signal processing, and was first introduced by Bo Hjorth \cite{15}.

\begin{table}
\caption{Feature Set}
\label{Feature Set}
\resizebox{\columnwidth}{!}{\begin {tabular}{|p{2.5cm}|p{4.1cm}|}
\hline
\textbf{TIME DOMAIN FEATURES} & \textbf{DESCRIPTION}                                                                                                \\ \hline
Mean                          & Mean value of the signal in a window.                                                                               \\ \hline
Standard Deviation            & Deviation from the mean value of the signal in a window.                                                            \\ \hline
Skewness                      & Measurement of lack of symmetry or asymmetry of an EEG  signal.                                                     \\ \hline
Kurtosis                      & Essentially measuring the complexity of EEG signal, it  determines if the signal has a peak or is flat at the mean. \\ \hline
Hjorth Activity               & Variance of EEG signal in a window.                                                                                 \\ \hline
Hjorth Mobility               & Measure of proportion of standard deviation of the power spectral density.                                          \\ \hline
Hjorth Complexity             & Compares similarity of the EEG signal to a pure sine wave.                                                          \\ \hline
\textbf{ENTROPY FEATURES}     & \textbf{DESCRIPTION}                                                                                                \\ \hline
Permutation Entropy           & Measure of the local complexity in a signal.                                                                        \\ \hline
Shannon Entropy               & Measure of uncertainty in random process or quantities.                                                             \\ \hline
Approximate Entropy           & Measure of the regularity and fluctuation in a time series.                                                         \\ \hline
Sample Entropy                & Improved form of Approximate entropy, described as index of regularity. It reduces the bias caused by self-matching. \\ \hline
\end{tabular}}
\end{table}

The aforementioned 11 features were calculated for each channel and this resulted in a large number of feature vectors, which added to the computation overhead owing to its huge dimensionality. Many attributes may be highly correlated and thus, redundant, adding to the high dimensionality of data. High-dimensional data pose a problem of overfitting to predictive based models. To tackle this problem, Principal Component Analysis (PCA) --- a statistical method that reduces dimensionality by projecting most relevant attributes into lower dimensional space was used in this study. This improves predictive performance of the models by eliminating less significant attributes. 

\begin{table*}[t]
\caption{ProtoNN Results}
\label{ProtoNN Results}
\centering
\setlength{\tabcolsep}{8pt}
\def\arraystretch{1.4}
\begin{tabular}{|c|c|c|c|c|c|c|c|c|c|}
\hline
\multirow{2}{*}{\textbf{\begin{tabular}[c]{@{}c@{}}WINDOW LENGTH\\ (IN SECONDS)\end{tabular}}} & \multirow{2}{*}{\textbf{PCA}} & \multirow{2}{*}{\textbf{\begin{tabular}[c]{@{}c@{}}TEST\\ ACCURACY\end{tabular}}} & \multirow{2}{*}{\textbf{\begin{tabular}[c]{@{}c@{}} MODEL SIZE \\ (IN BYTES)\end{tabular}}} & \multicolumn{2}{c|}{\textbf{PRECISION}} & \multicolumn{2}{c|}{\textbf{RECALL}} & \multicolumn{2}{c|}{\textbf{F1}}    \\ \cline{5-10} 
                                                                                               &                               &                                         &                                                 & \textbf{Class 0}   & \textbf{Class 1}   & \textbf{Class 0}  & \textbf{Class 1} & \textbf{Class 0} & \textbf{Class 1} \\ \hline
\multirow{4}{*}{1}                                                                             & 20                            & 0.70                                    & 1760                                            & 0.76               & 0.70               & 0.66              & 0.80             & 0.71             & 0.75             \\ \cline{2-10} 
                                                                                               & 50                            & 0.72                                    & 2960                                            & 0.8                & 0.75               & 0.73              & 0.82             & 0.76             & 0.78             \\ \cline{2-10} 
                                                                                               & 70                            & 0.75                                    & 3760                                            & 0.8                & 0.76               & 0.73              & 0.82             & 0.76             & 0.79             \\ \cline{2-10} 
                                                                                               & 100                           & 0.75                                    & 4960                                            & 0.81               & 0.8                & 0.79              & 0.82             & 0.80             & 0.81             \\ \hline
\multirow{4}{*}{2}                                                                             & 20                            & 0.71                                    & 1760                                            & 0.75               & 0.72               & 0.67              & 0.8              & 0.71             & 0.76             \\ \cline{2-10} 
                                                                                               & 50                            & 0.72                                    & 2960                                            & 0.79               & 0.78               & 0.73              & 0.82             & 0.76             & 0.80             \\ \cline{2-10} 
                                                                                               & 70                            & 0.73                                    & 3760                                            & 0.79               & 0.77               & 0.74              & 0.81             & 0.76             & 0.79             \\ \cline{2-10} 
                                                                                               & 100                           & 0.76                                    & 4960                                            & 0.82               & 0.82               & 0.80              & 0.83             & 0.81             & 0.82             \\ \hline
\multirow{4}{*}{4}                                                                             & 20                            & 0.76                                    & 1760                                            & 0.79               & 0.84               & 0.83              & 0.80             & 0.81             & 0.82             \\ \cline{2-10} 
                                                                                               & 50                            & 0.77                                    & 2960                                            & 0.80               & 0.84               & 0.83              & 0.81             & 0.81             & 0.82             \\ \cline{2-10} 
                                                                                               & 70                            & 0.75                                    & 3760                                            & 0.80               & 0.85               & 0.84              & 0.81             & 0.82             & 0.83             \\ \cline{2-10} 
                                                                                               & 100                           & 0.76                                    & 4960                                            & 0.83               & 0.86               & 0.86              & 0.84             & 0.84             & 0.85             \\ \hline
\multirow{4}{*}{8}                                                                             & 20                            & 0.75                                    & 1760                                            & 0.78               & 0.78               & 0.84              & 0.82             & 0.81             & 0.80             \\ \cline{2-10} 
                                                                                               & 50                            & 0.74                                    & 2960                                            & 0.79               & 0.84               & 0.85              & 0.82             & 0.82             & 0.83             \\ \cline{2-10} 
                                                                                               & 70                            & 0.75                                    & 3760                                            & 0.8                & 0.85               & 0.90              & 0.84             & 0.85             & 0.84             \\ \cline{2-10} 
                                                                                               & 100                           & 0.77                                    & 4960                                            & 0.78               & 0.85               & 0.86              & 0.83             & 0.82             & 0.84             \\ \hline
\multirow{4}{*}{16}                                                                            & 20                            & 0.78                                    & 1760                                            & 0.79               & 0.80               & 0.84              & 0.85             & 0.81             & 0.82             \\ \cline{2-10} 
                                                                                               & 50                            & 0.79                                    & 2960                                            & 0.81               & 0.85               & 0.88              & 0.83             & 0.84             & 0.84             \\ \cline{2-10} 
                                                                                               & 70                            & 0.79                                    & 3760                                            & 0.82               & 0.84               & 0.87              & 0.81             & 0.84             & 0.82             \\ \cline{2-10} 
                                                                                               & 100                           & 0.81                                    & 4960                                            & 0.84               & 0.85               & 0.86              & 0.87             & 0.85             & 0.86             \\ \hline
\end{tabular}
\end{table*}

In this research, the total features were reduced to subsets of 20, 50, 70 and 100 for experimentation.

\subsection{ML Subsystem}
\label{ML Subsystem}

The feature extractor is connected to a classification module which consists of a scalable, binary classifier called ProtoNN \cite{16}. ProtoNN models can be deployed on devices with scarce storages and constrained computational capacity. This kNN-based algorithm handles the trade-off between prediction accuracy and model size -- a solution not proposed in earlier literatures \cite{10}. ProtoNN implements this proposed solution by employing 3 key methods:
\begin{itemize}
    \item Projecting the entire data in low-dimension using a sparse projection matrix.
    \item Learning prototypes to represent the entire training dataset. This leads to flexibility and allows seamless generalization of ProtoNN.
    \item Learning the projection matrix jointly with the prototypes and their labels.
\end{itemize}

This allows the classifier to be deployed on devices that have RAM in the order of a few kilobytes. In contrast, kNN uses the entire training set for learning and prediction, hence is sizable for IoT devices. It also computes the distance of each test point to training point, making it slower for real-time prediction. 

\section{Results and Analyses}
\label{Results and Discussion}

In order to record inference time, all models were off-loaded to a Raspberry Pi 3 Model B whose specifications are provided in Table \ref{edge-device-specifications}.

\subsection{Metrics}
Assessing the performance of machine learning models can be done with the standard classification metrics which are extensively used in previous literature. In this study, the metrics used are listed below along with their mathematical formulae.

\begin{equation}
Accuracy =  \frac{TP+TN}{TP+TN+FP+FN} 
\end{equation}

\begin{equation}
Precision = \frac{TP}{TP+FP} 
\end{equation}

\begin{equation}
Recall = \frac{TP}{TP+FN} 
\end{equation}

\begin{equation}
F1 =  \frac{2*Precision*Recall}{Precision+Recall} = \frac{2*TP}{2*TP+FP+FN}
\end{equation}

$TP$ = Seizure events classified as seizure events

$FP$ = Non-seizure events classified as seizure events

$TN$ = Non-seizure events classified as non-seizure events

$FN$ = Seizure events classified as non-seizure events

\begin{table}[]
\caption{Edge-device specifications}
\label{edge-device-specifications}
\begin{tabular}{|l|l|}
\hline
SoC             & \begin{tabular}[c]{@{}l@{}}Broadcom BCM2837B0, Cortex-A53\\ (ARMv8) 64-bit\end{tabular} \\ \hline
RAM             & 1 GB                                                                                     \\ \hline
Operating Power & 5V / 2.5 A DC power unit                                                                \\ \hline
Clock Speed     & 1.4GHz                                                                                  \\ \hline
\end{tabular}
\end{table}

\vspace{1cm}

An additional metric known as Area Under receiver-operator-characteristics Curve \textit{(AUC)} is also used to measure the quality of predictive performance of the model. Other metrics used to evaluate the model performance on the embedded device are model size and inference time. Model size is the memory footprint of the classifier on the embedded device. Inference time is termed as the average time for the model to preprocess and predict the class of one segment of data.

\begin{table*}[]
\caption{kNN Results}
\label{kNN Results}
\setlength{\tabcolsep}{9.65pt}
\def\arraystretch{1.3}
\begin{tabular}{|c|c|c|c|c|c|c|c|c|c|}
\hline
\multirow{2}{*}{\textbf{\begin{tabular}[c]{@{}c@{}}WINDOW LENGTH\\ (IN SECONDS)\end{tabular}}} & \multirow{2}{*}{\textbf{PCA}} & \multirow{2}{*}{\textbf{K VALUE}} & \multirow{2}{*}{\textbf{\begin{tabular}[c]{@{}c@{}}TEST\\ ACCURACY\end{tabular}}} & \multicolumn{2}{c|}{\textbf{PRECISION}} & \multicolumn{2}{c|}{\textbf{RECALL}} & \multicolumn{2}{c|}{\textbf{F1}} \\ \cline{5-10} 
                                        &                               &                                   &                                    & \textbf{Class 0}         & \textbf{Class 1}         & \textbf{Class 0}        & \textbf{Class 1}       & \textbf{Class 0}      & \textbf{Class 1}     \\ \hline
\multirow{4}{*}{1}                      & 20                            & 37                                & 0.75                               & 0.66               & 0.84               & 0.81              & 0.71             & 0.72            & 0.77           \\ \cline{2-10} 
                                        & 50                            & 13                                & 0.78                               & 0.65               & 0.85               & 0.81              & 0.71             & 0.72            & 0.77           \\ \cline{2-10} 
                                        & 70                            & 37                                & 0.76                               & 0.70                & 0.83               & 0.80               & 0.73             & 0.74            & 0.77           \\ \cline{2-10} 
                                        & 100                           & 21                                & 0.778                              & 0.69               & 0.84               & 0.81              & 0.73             & 0.74            & 0.78           \\ \hline
\multirow{4}{*}{2}                      & 20                            & 16                                & 0.69                               & 0.71               & 0.68               & 0.69              & 0.70              & 0.70             & 0.69           \\ \cline{2-10} 
                                        & 50                            & 18                                & 0.75                               & 0.78               & 0.73               & 0.74              & 0.77             & 0.76            & 0.75           \\ \cline{2-10} 
                                        & 70                            & 15                                & 0.76                               & 0.73               & 0.80                & 0.78              & 0.74             & 0.75            & 0.77           \\ \cline{2-10} 
                                        & 100                           & 9                                 & 0.77                               & 0.75               & 0.78               & 0.77              & 0.76             & 0.76            & 0.77           \\ \hline
\multirow{4}{*}{4}                      & 20                            & 19                                & 0.78                               & 0.75               & 0.80                & 0.79              & 0.76             & 0.77            & 0.78           \\ \cline{2-10} 
                                        & 50                            & 8                                 & 0.78                               & 0.76               & 0.80                & 0.79              & 0.77             & 0.77            & 0.78           \\ \cline{2-10} 
                                        & 70                            & 11                                & 0.78                               & 0.76               & 0.81               & 0.80               & 0.77             & 0.78            & 0.79           \\ \cline{2-10} 
                                        & 100                           & 23                                & 0.81                               & 0.79               & 0.82               & 0.82              & 0.80              & 0.80             & 0.81           \\ \hline
\multirow{4}{*}{8}                      & 20                            & 15                                & 0.70                                & 0.66               & 0.72               & 0.70               & 0.68             & 0.68            & 0.70            \\ \cline{2-10} 
                                        & 50                            & 11                                & 0.80                                & 0.79               & 0.78               & 0.78              & 0.79             & 0.79            & 0.78           \\ \cline{2-10} 
                                        & 70                            & 22                                & 0.80                                & 0.77               & 0.79               & 0.79              & 0.78             & 0.78            & 0.78           \\ \cline{2-10} 
                                        & 100                           & 9                                 & 0.80                                & 0.75               & 0.81               & 0.80               & 0.76             & 0.77            & 0.79           \\ \hline
\multirow{4}{*}{16}                     & 20                            & 37                                & 0.69                               & 0.68               & 0.70                & 0.69              & 0.69             & 0.69            & 0.69           \\ \cline{2-10} 
                                        & 50                            & 5                                 & 0.78                               & 0.75               & 0.75               & 0.75              & 0.75             & 0.75            & 0.75           \\ \cline{2-10} 
                                        & 70                            & 3                                 & 0.78                               & 0.72               & 0.78               & 0.77              & 0.73             & 0.74            & 0.76           \\ \cline{2-10} 
                                        & 100                           & 6                                 & 0.78                               & 0.75               & 0.77               & 0.76              & 0.75             & 0.76            & 0.75           \\ \hline
\end{tabular}
\end{table*}

\subsection{ProtoNN Results and Analysis}

\begin{figure}
\centering
\includegraphics[scale = 0.4]{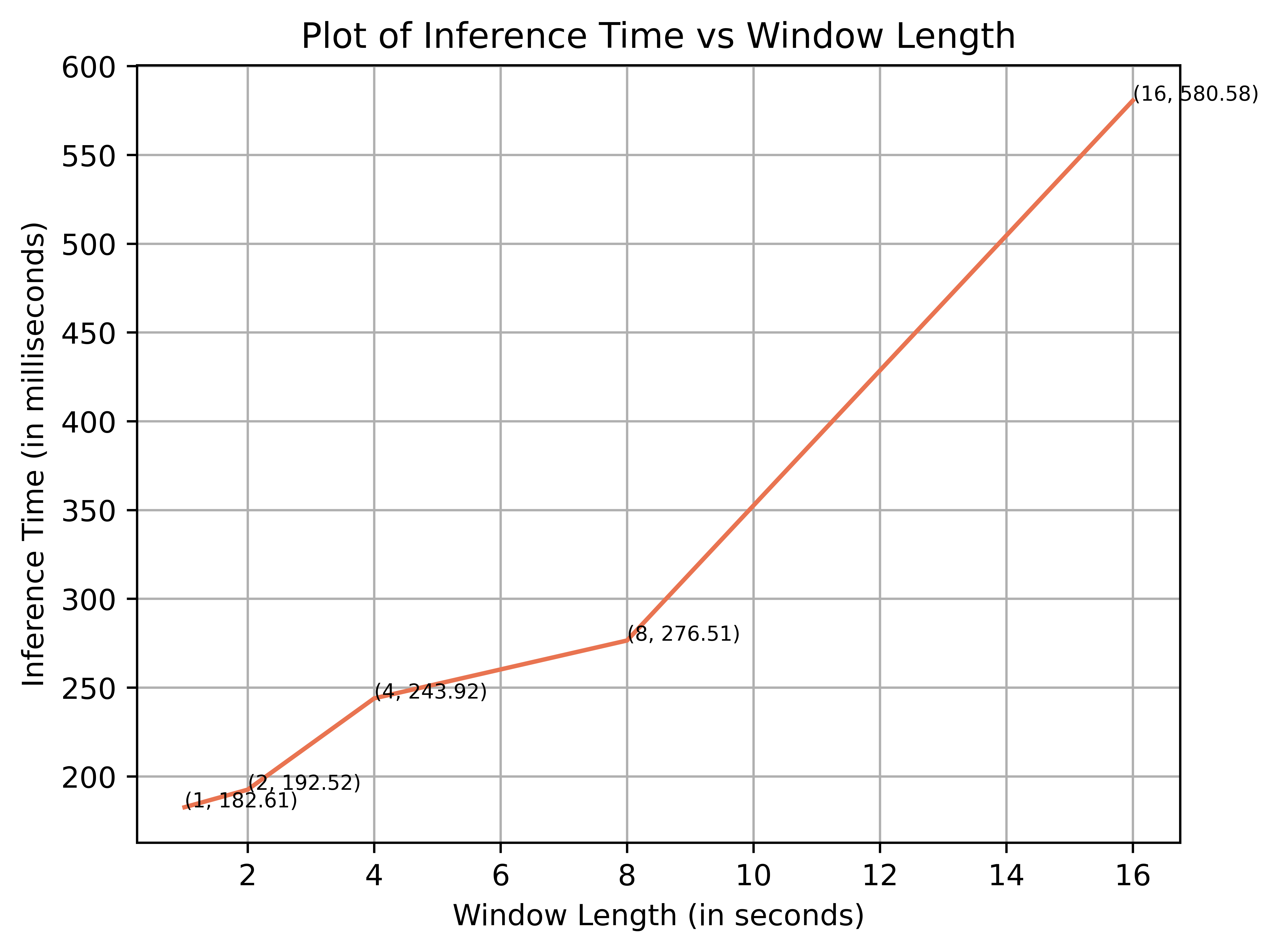}
\caption{Inference Time vs Window Length}
\label{inf_time_vs_win_len}
\end{figure}

\begin{figure}
\centering
\includegraphics[scale = 0.4]{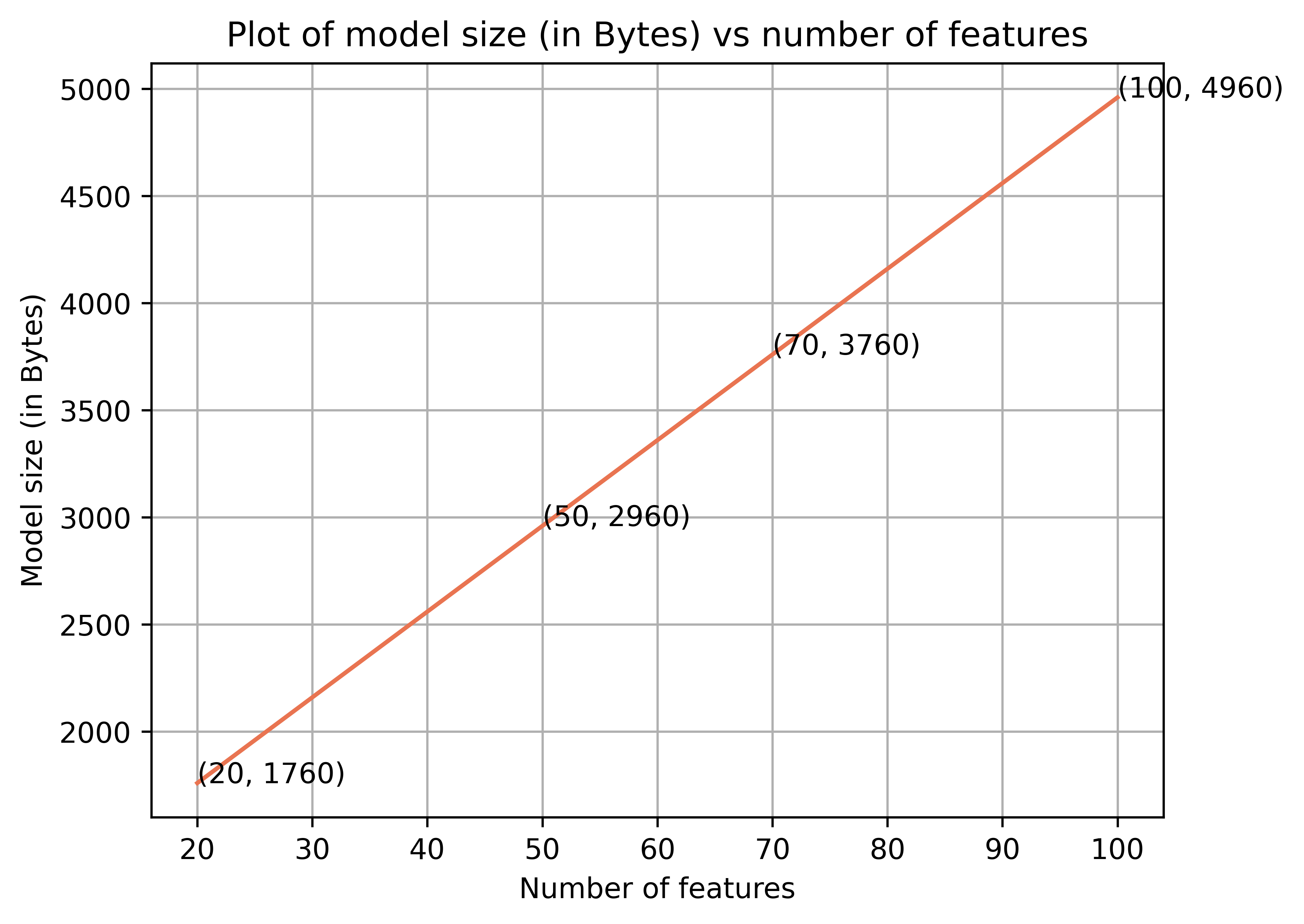}
\caption{Model Size vs No. of features}
\label{model_size_vs_nof}
\end{figure}

False negatives, in the case of seizure detection, are the most dangerous predictions as an undetected seizure leads to the neonate not receiving immediate attention and necessary treatment. To measure this parameter, the metric of sensitivity (recall of class 1) is closely observed. A model with high sensitivity produces lesser false negatives. From Table \ref{ProtoNN Results}, it can be discerned that the classifier trained on $w = 16$ seconds with 100 features delivers highest sensitivity of 87\%. Therefore, the optimal feature subset to be chosen for effective results in this study would be 100. As seen from Fig. \ref{inf_time_vs_win_len}, ProtoNN achieved least inference time for $w = 1$ second with 100 features. However, the best trade-off between sensitivity and inference time can be seen in $w = 4$ seconds where the sensitivity is 84\% and inference time is 243.92 milliseconds, demonstrating the quick response of the subsystem. Therefore, the optimal window length for this application is $w = 4$ seconds. From Fig. \ref{model_size_vs_nof}, as the number of features increase, the model size increases as well. However, the size of the subsystem is still in the range of a few kilobytes, with the heaviest model being just 4.84 KB. This substantiates the compactness of the model under resource-constrained settings. 

\subsection{Comparison with standard kNN Model}

To compare the performance of our classifiers with a baseline ML model, kNN was also tested on the same dataset, since the retrieval of nearest neighbours operation of ProtoNN is modelled on kNN. The kNN models were trained on all feature subsets, the results of which are tabulated in Table \ref{kNN Results}. Although kNN is comparable to ProtoNN in terms of accuracy, the small size and less prediction time of ProtoNN makes it more suitable for real-time NSD where deployability and rapid response is of paramount importance. The results therefore prove the superiority of ProtoNN on edge-devices compared to other models due to its scalability and feasibility. Currently, there exists no published research in the field of NSD with edge-constraints, thereby making our model the most compact one that can be used in quick and accurate seizure detection.

\section{Conclusion}

In this paper, a resource efficient, medically feasible approach for detecting neonatal seizure has been presented and compared with a baseline ML model. The proposed architecture makes an optimal trade-off between predictive score and inference time, thereby facilitating accurate and viable seizure detection with practical application. The low storage requirement of the model makes it abundantly suitable for deployability on edge-devices. In future, the system can be tweaked to perform seizure prediction earlier than its onset by using algorithms such as early stopping and recurrent neural networks. Further, this pipeline can be integrated into a wearable device that processes the EEG signals and makes real-time predictions.

\bibliographystyle{IEEEtran}
\footnotesize
\bibliography{IEEEabrv, citations}
\nocite{*}
\end{document}